\documentclass[aps,amsfonts,nofootinbib,article,twocolumn,showpacs,superscriptaddress]{revtex4}
\usepackage{graphicx}
\usepackage{amsmath}
\usepackage{amsfonts}
\usepackage{amssymb}

\def\openone{\leavevmode\hbox{\small1\kern-3.8pt\normalsize1}}
\def\N{\leavevmode\hbox{ Z \kern-8 pt\normalsize{Z}}}
\def\openone{\leavevmode\hbox{\small1\kern-3.8pt\normalsize1}}
\def\openJ{\leavevmode\hbox{J \kern-9.5pt\normalsize J}}
\def\openS{\leavevmode\hbox{ S \kern-9.3pt\normalsize S}}
\newcommand{\bb}{\begin{equation}}
\newcommand{\ee}{\end{equation}}
\newcommand{\eqb}{\begin{eqnarray}}
\newcommand{\eqf}{\end{eqnarray}}

\usepackage{color}

\begin{document}

\title{Quantization of the Universe as a Dirac Particle: Geometrical Unification of Gravitation, Dark Energy and Multiverses}

\author{Sergio A. Hojman}
\email{sergio.hojman@uai.cl}
\affiliation{UAI Physics Center, 
Universidad Adolfo Ib\'a\~nez, Santiago, Chile.}
\affiliation{Departamento de Ciencias, Facultad de Artes Liberales,
Universidad Adolfo Ib\'a\~nez, Santiago, Chile.}
\affiliation{Departamento de F\'{\i}sica, Facultad de Ciencias, Universidad de Chile,
Santiago, Chile.}
\affiliation{Centro de Recursos Educativos Avanzados,
CREA, Santiago, Chile.}

\author{Felipe A. Asenjo}
\email{felipe.asenjo@uai.cl}
\affiliation{UAI Physics Center,
Universidad Adolfo Ib\'a\~nez, Santiago, Chile.}
\affiliation{Facultad de Ingenier\'{\i}a y Ciencias,
Universidad Adolfo Ib\'a\~nez, Santiago, Chile.}

\author{Carlos A. Rubio}
\email{carubiof@uc.cl}
\affiliation{Facultad de F\'isica, Pontificia Universidad Cat\'olica de Chile,
Santiago, Chile.}

\begin{abstract}
A Friedmann--Robertson--Walker Universe is studied with a dark energy component represented by a quintessence field. The Lagrangian  for this system, hereafter called the Friedmann--Robertson--Walker--quintessence (FRWq) system, is presented.  It is shown that the classical Lagrangian reproduces the usual
two (second order) dynamical equations for the radius of the Universe and for the quintessence scalar field as well as a (first order)
constraint equation.
 It is noted that the quintessence field (dark energy) is related with the tlaplon field (appearing on dynamical theories of spacetime torsion). Our approach naturally unifies gravity and dark energy, as it is obtained that the Lagrangian and the equations of motion are those of a relativistic particle moving on a
two dimensional conformally flat spacetime, whereas the conformal metric factor is related to the dark energy scalar field potential.
We proceed to quantize the system in three different schemes. First, we assume the Universe as a spinless particle  (as it is common in literature), obtaining a quantum theory for a Universe described by the Klein--Gordon equation.
Second, we push the quantization scheme further assuming the Universe as a Dirac particle, and therefore constructing its corresponding Dirac and the Majorana theories.
With the different theories we calculate the expected values for the scale factor of the Universe. They depend on the type of quantization scheme used. The differences between the Dirac and Majorana schemes are highlighted.
The implications of the different quantization procedures are discussed. Finally the possible consequences for a multiverse theory of the Dirac and Majorana quantized Universe are briefly considered.
\end{abstract}

\pacs{}

\maketitle
\section{Introduction}

Quintessence is the name of one model put forward in order to explain
that the rate of expansion of the Universe increases. The model modifies the equations of General Relativity by adding a
Lagrangian density for a massless scalar (quintessence) field rolling down a potential minimally coupled
to the usual Einstein--Hilbert Lagrangian density \cite{tsu, copeland}.

The equations of motion for the Friedmann--Robertson--Walker--quintessence (FRWq) system are obtained from Einstein's equations modified by the addition of the quintessence field.  By symmetry considerations, the equations are simplified, which transforms them
from partial to ordinary differential equations. They consist of a set of two ordinary dynamical second order differential equations which govern the evolution
of the dynamical variables (the radius of the universe and the scalar quintessence field) and one ordinary first order differential equation
which constraints the initial values and velocities of the dynamical variables.

This system has been studied extensively both in classical \cite{tsu, copeland,capozz} as well as in quantum \cite{bojo,Steinhardt} cosmologies. Its importance is justified as this model is used to give an explanation to dark energy.
Several of the articles which deal with the subject, write down a Lagrangian formulation for the FRWq system.
Nevertheless, to the best of our knowledge, all of the work published up to now is based on a classical Lagrangian
which gives rise to the two dynamical equations but does not yield the contraint equation \cite{sahfaa}.

Along this work, we propose a model for a FRWq system that can be studied as a relativistic pointlike particle. This will allow us to quantize the system  creating different quantum cosmological models.
The first step in this model consists in constructing a new Lagrangian without the previously mentioned problem.
The Lagrangian presented in this article gives rise to
two dynamical equations and one constraint equation \cite{sahfaa}.
We show that the FRWq system may be completely described in terms of a Lagrangian similar to that of a relativistic pointlike particle
moving on a two dimensional conformally flat gravitational field. This is presented in Secs. II and III.
The two dimensional conformal factor is
a function of the radius of the universe and of the scalar quintessence field which naturally play
the role of coordinates in this two dimensional (mini) superspace. Hence, the relativistic particle description of the FRWq system is only possible with the merging of spacetime and quintessence. Importantly enough, a similarity of the quintessence field with the tlaplon field is pointed out in Sec. IV. The tlaplon field is  responsible for a dynamical (propagating) torsion of the spacetime \cite{hrrs}. Therefore, a link between torsion and dark energy is hinted.

The previous concepts and identifications between the classical cosmological spacetime and the  quintessence field will lead us to a straightforward construction of a quantum theory in Sec. V. Notice that this is only possible because the unification achieved through the relativistic particle description. Quantizing the model as a spinless particle will give rise to a Klein-Gordon theory which is a generalization of  the Wheeler–DeWitt equation. Also, following Breit's prescription \cite{breit} we can quantize the model in a way analogous  to that of a spin particle, producing Dirac and  Majorana theories for the Universe.

Finally, in Sec. VI, different aspects of our models are discussed. The validity of the Dirac quantization scheme is highlighted in that section showing how the same theory can be obtained using different approaches. Also, a connection of this theory with Multiverses is mentioned.

\section{Lagrangian for FRWq system}

In general, it is possible to write a Lagrangian density $\mathcal{L}$ for the evolution of the
spacetime metric $g_{\mu \nu}(x^\alpha)$ in interaction with a
massless scalar  field $\phi(x^\beta)$ which may be identified with the quintessence field. The general Lagrangian  is
\begin{equation}
\mathcal{L}=\sqrt{-g}\left(\frac{R-2\Lambda}{2{\cal G}}+{\mathcal L}_\phi\right)\, ,
\label{lFRWq}
\end{equation}
where $g$ stands for the determinant of the metric $g_{\mu\nu}$, ${\cal G}= 8\pi G/c^4$ (with $G$ as the gravitational constant and $c$ the speed of light), $\Lambda$ is the  cosmological constant, and $\mathcal{L}_\phi$ is the Lagrangian density for the massless scalar field
\begin{equation}
\mathcal{L}_\phi=\epsilon\left(\frac{1}{2}g^{\mu \nu} \phi,_\mu
\phi,_\nu - {\cal V}(\phi)\right)\, ,\label{lFRWq2}
\end{equation}
where ${\cal V}(\phi)$ is  (up to now) an unspecified potential for the
scalar field $\phi$, whereas $\epsilon$ is a parameter that classifies the nature of the scalar field:  $\epsilon=1$ yields the Lagrangian density for usual scalar fields while $\epsilon=-1$ defines the gravitational Lagrangian modified with the quintessence field \cite{Steinhardt}.

 As it is well known, the Lagrangian density $\mathcal{L}$ defined by \eqref{lFRWq} is
 singular. The action $S$
\begin{equation}
S= \int \mathcal{L}\ d^4 x \label{action}\, ,
\end{equation}
gives rise (upon variation with respect to the metric tensor $g_{\mu
\nu}$) to gauge invariant (generally covariant) and constrained
 Einstein field equations coupled to matter, i.e.,
 \begin{equation}
G_{\mu \nu}+ \Lambda g_{\mu\nu} ={\cal G}\, T_{\mu \nu}\, , \label{ee}
\end{equation}
where $G^{\mu \nu}$ is the Einstein tensor and $T^{\mu \nu}$ is the
energy--momentum tensor of matter
\begin{equation}\label{}
  T_{\mu\nu}=g_{\mu\nu} \mathcal {L}_\phi-2\frac{\delta \mathcal {L}_\phi}{\delta g^{\mu\nu}}\, .
\end{equation}
Variation of the action $S$ with respect to $\phi$ yields the Klein--Gordon equation for the massless scalar field
 \begin{equation}
\Box \phi + \frac{d {\cal V}(\phi)}{d \phi}= 0\, , \label{kge}
\end{equation}
where $\Box$ is the d'Alembert operator in curved spacetimes.

In order to study a cosmological model with quintessence, let us take the line element for an isotropic and
homogeneous Friedmann--Robertson--Walker (FRW) spacetime, with the metric defined by \cite{ryden}
\begin{equation}
ds^2=-dt^2+ a(t)^2\left[\frac{d\rho^2}{1-k\rho^2} +\rho^2(d\theta^2+
\sin^2\theta d\phi^2)\right]\, , \label{frwm}
\end{equation}
where $a(t)$ is related to the radius of the Universe.
 The number $k$ is called the curvature constant and it can take the values $k=-1, 0, 1$  measuring the negative, zero or positive curvature of the Universe.

When Einstein equations coupled to matter are written in terms of
the line element \eqref{frwm}, and under the assumption that the scalar field $\phi$ depends on time only , we get two second order dynamical
and one first order constraint. Setting ${\cal G}=1$, the dynamical equations read
\begin{equation}
2\frac{\ddot a}{a}+\left(\frac{\dot a}{a}\right)^2 +\
\frac{k}{a^2} -\epsilon\left(\frac{1}{2}\ \dot \phi^2-V(\phi)
\right)= 0, \label{FRWqe1}
\end{equation}
and
\begin{equation}
\ddot\phi + 3\frac{\dot a}{a} \dot\phi + \frac{d
V(\phi)}{d\phi}=0\, , \label{FRWqe2}
\end{equation}
where we have introduced the new potential $V(\phi)={\cal V}(\phi)-\epsilon \Lambda$. The constraint equation is
\begin{equation}
3\left(\frac{\dot a}{a}\right)^2 +3\frac{k}{a^2}+\epsilon\left(\frac{1}{2}\dot \phi^2+ V(\phi) \right) =0\, . \label{FRWqc}
\end{equation}
Another useful equation can be obtained manipulating Eqs. \eqref{FRWqe1} and \eqref{FRWqc}
\begin{equation}\label{usefulEC}
  \frac{\ddot a}{a}=\frac{\epsilon}{3}\left(\dot\phi^2-V\right)\, .
\end{equation}

For $\epsilon=-1$, the set \eqref{FRWqe1}-\eqref{usefulEC} becomes the FRWq system. This set of equations has been already studied  and solved for quintessence by Capozziello and Roshan \cite{capozz} for different scenarios and configurations of matter.

In this article, we explore the analogy of this system with a relativistic particle \cite{sahfaa} which implies, as we will show, a geometric unification of gravity and quintessence which may be interpreted as a torsion potential field, one form of dark energy as it will be argued below. With this purpose in mind, we examine a Lagrangian $L$ that gives
rise to the dynamical equations \eqref{FRWqe1} and \eqref{FRWqe2}. This Lagrangian is
\begin{equation}
L= 3 a {\dot a}^2 - 3 k a +\epsilon a^3\left(\frac{1}{2}\dot\phi^2-V(\phi) \right)\, . \label{lag1}
\end{equation}
However, it is important to emphasize that the Lagrangian \eqref{lag1} does
not produce the constraint equation \eqref{FRWqc}. This constraint is equivalent to
imposing that the Hamiltonian $H$ associated to $L$ vanishes, i.e.,
\begin{equation}
H \equiv \frac{\partial L}{\partial \dot a} \dot a + \frac{\partial
L}{\partial \dot \phi} \dot \phi-  L=0\, . \label{H0}
\end{equation}
It is a remarkable fact that the change of variables
\begin{equation}
r={\frac{2\sqrt 6}{3}}\ a^{3/2}\, ,\qquad
\theta=\frac{3}{2\sqrt 6}\ \phi\, ,\label{theta}
\end{equation}
recasts the Lagrangian $L$ in a ``kinetic energy minus potential energy ($T-V$)'' form (disregarding the fact that both $T$ and $V$ have the ``wrong signs''
in the $\theta$ associated terms, for the quintessence case)
\begin{equation}
L\rightarrow{\bar L}=\frac{1}{2}(\dot r^2+\epsilon r^2\dot \theta^2)-\bar V(r,\theta)\, ,\label{lag2}
\end{equation}
where $\bar V(r, \theta)$ is a general  potential
\begin{equation}
\bar V(r,\theta)=3 \left(\frac{3}{8}\right)^{\frac{1}{3}}k\, r^{2/3}
+\epsilon\frac{3}{8}r^2 V(\theta)\, .\label{pot1}
\end{equation}

Note that in the Lagrangian ${\bar L}$ defined in \eqref{lag2}, which
describes the evolution of a FRWq Universe  in the presence of geometry (represented either by $r$ or by $a$) and dark energy (represented by
$\epsilon=-1$, and either $\theta$ or $\phi$), shows that the Universe evolves as a
relativistic particle moving on a two dimensional surface under the influence
of the potential \eqref{pot1}, thus geometrically unifying gravity and dark energy.

Nevertheless, the Lagrangian \eqref{lag2} does not produce the
constraint equation \eqref{FRWqc}. In order to construct a
Lagrangian which in fact gives rise to all three equations
\eqref{FRWqe1}, \eqref{FRWqe2} and \eqref{FRWqc} it is enough to
recall that \cite{h2,sahfaa} Jacobi--Maupertuis and Fermat principles give rise to
identical equations of motion in classical mechanics and geometrical
(ray) optics except for the fact that Fermat principle also produces
a constraint equation. It is worth mentioning that exactly the same
results are reached for the case of a relativistic particle moving on a two
dimensional conformally flat spacetime.

\section{Quintessence and Fermat-like Lagrangian}

From now on, we just focus in the case of $\epsilon=-1$ which describes the dark energy (quintessence) scenario \cite{Steinhardt}.
The description of the FRWq system in terms of a Fermat--type Lagrangian is established by defining the relation between
the potential $\bar V(r, \theta)$ and the (``refraction index'') conformal factor $n(r, \theta)$
\begin{equation}
\bar V(r, \theta) \equiv -\frac{1}{2} \left[n(r, \theta)\right]^2\, ,
\label{Vn}
\end{equation}
and the constraint
\begin{equation}
\bar H \equiv \frac{\partial \bar L}{\partial \dot r} \dot r +
\frac{\partial \bar L}{\partial \dot \theta} \dot \theta- \bar L=0\, .
\label{E0}
\end{equation}

The Fermat--like Lagrangian $L_F$ which gives rise to all three equations
\eqref{FRWqe1}, \eqref{FRWqe2} and \eqref{FRWqc} is
\begin{equation}
L_F=  n(r,\theta) \sqrt{\left(\frac{d r}{d\lambda}\right)^2-r^2 \left(\frac{d
\theta}{d\lambda}\right)^2}\, ,\label{LF}
\end{equation}
where $\lambda$ is,
in principle, an arbitrary parameter and $n(r,\theta) =   \sqrt{-2 \bar V(r, \theta)}$,
with $\bar V(r, \theta)$ defined in \eqref{pot1} as
\begin{equation}
\bar V(r,\theta)=3 \left(\frac{3}{8}\right)^{\frac{1}{3}}k\, r^{2/3}
-\frac{3}{8}r^2 V(\theta)\, ,\label{pot1Quint}
\end{equation}
with $V(\theta)={\cal V}(\theta)+\Lambda$. Thus, the Lagrangian \eqref{LF} may be appropriately rewritten as
\begin{equation}
L_F=  \sqrt{-2 \bar V(r, \theta)\left[ \left(\frac{d r}{d\lambda}\right)^2-r^2 \left(\frac{d
\theta}{d\lambda}\right)^2\right]}\, .\label{LF1}
\end{equation}

To reproduce the relativistic equations of motion, $\lambda$ is defined by L\"{u}neburg's parameter
choice \cite{lune,h2}
\begin{equation}
 \sqrt{\left(\frac{d r}{d\lambda}\right)^2-r^2 \left(\frac{d
\theta}{d\lambda}\right)^2} =  n(r,\theta)\, .\label{lune}
\end{equation}

It is straightforward to prove that varying the Lagrangian \eqref{LF1} with respect to $r$ and $\theta$ one gets Eqs. \eqref{FRWqe1}, \eqref{FRWqe2} (when rewritten in terms of $r$ and $\theta$) while the constraint \eqref{FRWqc} arises because the Hamiltonian associated to a Lagrangian which is a homogeneous function of degree one in the velocities vanishes identically, much in the same way it happens in the Lagrangian description of the dynamics of a relativistic particle. This statement may equivalently be related to the eikonal equation in geometrical optics.

\section{Relation with Einstein Cartan theory and dynamical torsion}
\label{dytorsion}

Consider Einstein--Cartan (EC) theory, i.e., a gravitational theory which generalizes General Relativity by taking into account both curvature and torsion of spacetime. In EC theory, the connection coefficients are constructed as the addition of the (symmetric) Christoffel symbols constructed out of the spacetime metric $g_{\mu \nu}$ and an antisymmetric part which is constructed using the torsion tensor. The EC theory is constructed in such a way that General Relativity is recovered when torsion vanishes \cite{hehl}.

The coupling of the gravitational field (including torsion) to the electromagnetic field may now proceed using the conventional minimal coupling (combined with gauge invariance) scheme which implies that the electromagnetic field couples to the (metric defined symmetric) Christoffel symbols portion of the connection only. It does not couple to the (antisymmetric) torsion part of the connection \cite{hehl}.

In 1978, the tlaplon strategy was devised as an attempt to couple electromagnetism to torsion while preserving gauge invariance and minimal coupling. The tlaplon
scheme consists in writing the (24 component) torsion field as the gradient of a single scalar field, the tlaplon, allowing in
the process to couple photons to torsion while introducing slight modifications of the minimal coupling and gauge invariance schemes. These modifications have the appropriate behavior for the case of vanishing torsion.

Nevertheless, if the strict conventional definitions of gauge invariance and minimal coupling are enforced, then electromagnetic fields do not couple to torsion \cite{hehl}, in other words, torsion is one form of dark energy.

The EC Lagrangian for a dynamical spacetime with torsion written for the tlaplon case, i.e., for the case in which torsion depends on one scalar dynamical field only, is identical to the quintessence Lagrangian (up to the the addition of a cosmological constant term). Further specialization to the FRW metric yields the Friedmann Robertson Walker quintessence model (FRWq).

Therefore, the FRWq model may equivalently be thought of as a solution of the tlaplon--like EC theory, where torsion, represented by a single scalar field (the tlaplon), constitutes dark energy which neither couples to nor produces electromagnetic fields.

When quintessence is considered, $\epsilon=-1$, then the Lagrangian \eqref{lFRWq} is completely equivalent to the one defined for the tlaplon field \cite{hrrs} up to the addition of the cosmological constant term. The tlaplon (scalar) field was introduced to couple torsion with electromagnetic fields, however this feature is not essential to have dynamical torsion (i.e. one can assume that the coupling parameter is extremely small since this interaction has been not measured yet).  The torsion is thus  generated by the gradient of the tlaplon field \cite{hrrs}. This is in contrast to the Einstein--Cartan theory where the torsion is a non-propagating field (algebraically related to the spin of matter).

\section{Quantization}
\label{quantizationquantization}

We have showed that the FRWq dynamics may be described completely in terms of a Lagrangian similar to that of a relativistic particle. We can now go further in our scheme and proceed to quantize this model described by FRWq Lagrangian.

The quantization of cosmological Universes has been a prosperous field for decades.  This field, called quantum cosmology, attemps to construct a quantum theory for the entire Universe. However there is not a unique form to achieve this. Even more there is no certainty about the correct procedure to follow. Several possibilities have been carried out (for a comprehensive review, we refer the reader to Ref.~\cite{bojo}).  The most famous procedure correspond to perform a canonical quantization of the classical dynamical equations for the evolution of the FRW Universe. This is performed by replacing the momentum by a derivative operator on the scale factor variable. The final equation is widely  and generically known as the Wheeler–DeWitt equation \cite{dewitt,hawk1}. This equation will depend on the main features we want to study of the Universe. Thus, quantum cosmologies have been studied for Universes with dynamical vacuum  in de Sitter cosmologies \cite{alvarenga}, in anti–-de Sitter spacetimes \cite{oliveraG}, in radiation-dominated Universe \cite{lemos}, in Universes with cosmological constant \cite{mone, mone2}, in $f(R)$ gravity \cite{vaki, huang}, in conformal theory \cite{barbour}, and with a massive scalar field \cite{hawk2},  among plenty of other works.

Below we will proceed quantizing in three different ways. First, we will use canonical quantization of the classical Lagrangian  \eqref{LF1} modelling the Universe as a relativistic particle (producing a Klein-Gordon equation). Secondly, we quantize the FRWq Universe as a relativistic Dirac particle (Dirac or Majorana theories) given a proper physical justification for this procedure.

First, let us rewrite the Lagrangian \eqref{LF1} as
\begin{equation}
L_F=  \sqrt{\bar V(\xi, \theta)\,  e^{2\xi}\left( \dot\theta^2- \dot\xi^2\right)}\, ,\label{LF2}
\end{equation}
where we have introduced the new variable $\xi=\ln r$, and then $\bar V\equiv\bar V(\xi, \theta)=\bar V(r, \theta)$ and $\dot\theta={d\theta}/{d\lambda}$, $\dot\xi={d\xi}/{d\lambda}$.
The previous Lagrangian coincides with the one for a relativistic particle on a  two-dimensional conformally flat spacetime. The corresponding conformal flat metric is
\begin{equation}\label{}
g_{00}=\Omega^2\, ,\, \, g^{00}=\frac{1}{\Omega^2}\, ,\, \,g_{11}=-\Omega^2\, ,\, \, g^{11}=-\frac{1}{\Omega^2}\, ,
\end{equation}
where
\begin{equation}\label{functionOmegaa}
\Omega\equiv\sqrt{\bar V}e^\xi\, ,
\end{equation}
such that the Lagrangian \eqref{LF2} is written as $L_F=  \sqrt{g_{\mu\nu}dx^\mu dx^\nu/d\lambda^2}$, with $x^0=\theta$ and $x^1=\xi$.

In order to avoid problems with the procedure of canonical quantization of the FRWq system, we restrict ourselves to the cases where $\bar V >0$, and to consider static manifolds  only \cite{saa}, where there exists a family of spacelike surfaces which are always orthogonal to a timelike Killing vector. This implies that ${\partial_\theta g_{\mu\nu}}=0$, or
\begin{equation}\label{thetaindepv}
\frac{\partial\bar V}{\partial\theta}=0\, ,
\end{equation}
which means that the original potential $V(\theta)$ is a constant. Thereby, for the current quantization process, $V(\theta)$ is essentially equal to a cosmological constant.

We have just restricted ourselves to the cases in which the FRWq Lagrangian is $\theta$ independent. The associated Noether conservation law insures that $\dot\theta$ does not change
sign (see, for instance \eqref{FRWqe2} for a $\phi$ independent potential) and, therefore, $\theta$ may be used as the evolution (time) variable which is exactly what the variational principle and quantization procedure suggest.
Therefore, notice that the quintessence field acts as a super-time in this new description where the particle is moving in an ``effective'' two-dimensional conformally flat spacetime.

Classically, it can be rigorously shown \cite{hanson} that the Hamiltonian for the system described by Lagrangian \eqref{LF2} is
\begin{equation}\label{Hclass1}
  H=\sqrt{g_{00}}\sqrt{1-g^{11}\pi^2}=\sqrt{g_{00}}\sqrt{1+\frac{\pi^2}{\Omega^{2}}}\, ,
\end{equation}
where $\pi$ is the canonical momentum.
We use this Hamiltonian to construct the quantum theory for the FRWq system.  In order to avoid factor ordering issues, the quantum Hamiltonian operator $\mathcal H$ may be constructed from its classical analogue \eqref{Hclass1} as
\begin{equation}\label{momentumoperator1}
  \mathcal H=\Omega^{1/2}\sqrt{1+\hat p^2}\, \Omega^{1/2}\, ,
\end{equation}
where $\hat p$ is the momentum operator defined as
\begin{equation}\label{operDD}
  \hat p=\sqrt{-g^{11}}\hat\pi=\frac{\hat\pi}{\Omega}=-\frac{i}{\Omega}\frac{\partial}{\partial\xi}\, ,
\end{equation}
because of $\hat\pi=-i{\partial}_\xi$. In this way, the quantum equation that describe the quantization of the FRWq system is
\begin{equation}\label{quantumeq}
  i\hbar\frac{\partial\Psi}{\partial\theta}=\mathcal H\Psi\, ,
\end{equation}
where $\Psi$ is the wavefunction of the FRWq system.

In principle, one may ask whether there are ways to construct other Hamiltonian operators that differ from \eqref{momentumoperator1}, giving rise to quantum theories which are not equivalent to the one described by Eq.~\eqref{quantumeq} (see, for instance \cite{hojmanmontemayor}). This point is subtle, and the answer is affirmative, however the operator \eqref{momentumoperator1} has the advantage that it reproduces results from quantum field theory in curved spacetimes as we will see below.

In what follows we proceed to quantize the FRWq theory in two mayor different ways. In the first case, we quantize the system  using a procedure devised for spinless particles. This approach produces a Klein-Gordon equation for the wavefunction of the FRWq Universe. The second case corresponds to the quantization of the FRWq system as a Dirac particle. An argument which leads us to follow this path of quantization will be given below.

\subsection{Quantization of the FRWq system as a Klein-Gordon particle}

One way to canonically quantize the relativistic spinless particle can be obtained following the method developed by Gavrilov and Gitman \cite{gavrilov}. This procedure is a consistent way to construct the quantum theory along  Dirac's theory for gauge and constrained systems \cite{diracT, Teitel}.

We will not reproduce the calculations for this quantization scheme here, but we limit ourselves to exhibit the results of applying this method. The quantization for the FRWq system produces the quantum equation \cite{gavrilov}
\begin{equation}\label{KG1G}
  i\partial_\theta\Psi=\hat h \Psi\, ,
\end{equation}
(taking $\hbar=1$ for convenience) where $\Psi$ is the spinor
\begin{equation}\label{}
  \Psi=\left(
         \begin{array}{c}
           \chi \\
           \psi \\
         \end{array}
       \right)\, ,
       \end{equation}
  and $\hat h$ is a matrix Hamiltonian
              \begin{equation}
       \hat h=\left(
                \begin{array}{cc}
                  0 & -\partial_\xi^2+ \Omega^{2} \\
                  1 & 0 \\
                \end{array}
              \right)\, .
  \end{equation}

Let us notice that these are not two dynamical equations for a spinor, but Eq.~\eqref{KG1G} will produce the constraint $i\partial_\theta\psi=\chi$. Therefore, Eq.~\eqref{KG1G} gives rise to a  Klein-Gordon equation
\begin{equation}\label{KGfinalG}
  0=\frac{\partial^2\psi}{\partial \theta^2}-\frac{\partial^2\psi}{\partial \xi^2}+\Omega^2\psi\, .
\end{equation}

The wavefunction $\psi$ represents the probability amplitude  obtained by using the quantization of the FRWq Universe system as a Klein-Gordon particle. It obeys the same equation of the Klein-Gordon field in Minkowski space but now with an effective mass term whose origin is the conformal metric. A similar result is obtained when a scalar field is quantized in an expanding curved spacetime background \cite{muk} obtaining a mass-corrected term due to a conformal time. However we must emphasize that our treatment is different because in our approach it is the spacetime itself what is quantized.

The Klein-Gordon probability density is not positive definitive. For the case at hand, one can calculate from Eq.~\eqref{KGfinalG} the probability density as
\begin{equation}\label{}
  |\psi(\theta)|^2=\int_0^{a_f} \frac{da}{i a}\left(\psi^*\frac{\partial\psi}{\partial\theta}-\frac{\partial\psi^*}{\partial\theta}\psi\right)\, ,
\end{equation}
where $\psi^*$ is the complex conjugated of the wavefunction \eqref{expanKG}, and $a_f$ is some value of the scale factor that can be equal or larger than its present value. We notice that the probability density depends on $\theta$. In the same way, for our case, the expected value of the scale factor is \cite{strange}
\begin{equation}\label{valoraKGspec}
\left\langle a(\theta)\right\rangle=\frac{1}{|\psi|^2}\int_0^{a_f} \frac{da}{i} \left(\psi^* \frac{\partial\psi}{\partial\theta}-\frac{\partial\psi^*}{\partial\theta}\psi\right)\, ,
\end{equation}
which depends on $\theta$ as well.

 We can solve Eq.~\eqref{KGfinalG} exactly for $k=0$ and any constant $V$. Assuming the dependence $\psi(\xi)=\phi(\xi)\, e^{iE\theta}$, for a constant parameter  $E$, the previous equation becomes
\begin{equation}\label{KGquantumechanicalEq}
  \frac{\partial^2\phi}{\partial \xi^2}=\left(\Omega^2-E^2\right)\phi\, .
\end{equation}
For  spatially flat Universe $k=0$, where $\Omega^2=-3 V e^{4\xi}/8$ (being $V=V(\theta)$ a constant related to the cosmological constant),  the general solution is found to be
\begin{eqnarray}\label{}
  \phi(\xi)&=&C_1 J_{-iE/2}\left(\frac{e^{2\xi}}{4}\sqrt{\frac{3V}{2}}\right)\Gamma\left(1-\frac{iE}{2}\right)\nonumber\\
  &+&C_2 J_{iE/2}\left(\frac{e^{2\xi}}{4}\sqrt{\frac{3V}{2}}\right)\Gamma\left(1+\frac{iE}{2}\right)\, ,
\end{eqnarray}
where $\Gamma$ is the Euler gamma function, and $J_n$ is the modified Bessel function of the first kind of order $n$. By appropiate choice of constants $C_1$ and $C_2$, the wavefunction $\phi$ can be real.

To find solutions for $k\neq 0$ for the wavefunction $\psi$, we proceed assuming that the wavefunction can be decomposed as
\begin{equation}\label{expanKG}
\psi(\xi,\theta)=\psi(a,\theta)=\sum_{n=0}^\infty \psi_n(a)\, e^{i E_n\theta}\, ,
\end{equation}
where the reason to go back to the $a$-representation in the variables will clear in the following. Here $E_n$ is a parameter that, as we will see below, it can asociated to the energy of the $n$-state of the Klein-Gordon field $\psi_n$. Using this decomposition, we can find a suitable equation to be solved. First, let us define the field $u_n(a)=\sqrt{a}\, \psi_n(a)$. This new field follows the differential equation
\begin{equation}\label{KGquantumechanicalEq}
 -\frac{d^2 u_n}{d a^2}+V_{\mbox{\footnotesize{eff}}}(a)\,  u_n=\frac{9E_n^2}{4a^2} u_n\, ,
\end{equation}
where we have introduced the effective potential
\begin{equation}
V_{\mbox{\footnotesize{eff}}}(a)=\frac{9\Omega^2-1}{4a^2}\, ,
\end{equation}
that depends on the value of $k$.
Eq.~\eqref{KGquantumechanicalEq} can be converted in the following Ricatti equation
\begin{equation}
-i \frac{d\mu_n}{da}+\mu_n^2+\left(V_{\mbox{\footnotesize{eff}}}-\frac{9E_n^2}{4a^2}\right)=0\, ,
\end{equation}
where
\begin{equation}
\mu_n(a)=- i \frac{d}{da}\ln u_n\, .
\end{equation}
The above Ricatti equation can be general solved if a particular solution can be found for any $k$.

This is not a trivial task. An approximated solution of Eq.~\eqref{KGquantumechanicalEq} for the field $u_n$ can be obtained using
the Spectral Method \cite{spectral1,spectral2} in the expansion \eqref{expanKG}. This method is usually utilized in similar quantum theories for the Universe \cite{mone2}. As the wavefunction of the Universe must vanish at the origin as well as in infinity, the Spectral Method (SM) use the approximation that the wavefunction vanishes at some length $L$ (as large as we require). Thus, the SM allows us to expand the wavefunction $u_n$ in a Fourier series
\begin{equation}\label{SmKG}
  u_n(a)\approx \sum_{m=1}^N A_m^{(n)} \sin\left(\frac{m\pi}{L}a\right)\, ,
\end{equation}
where $A_m^{(n)}$ are constant coefficients that depends on the $n$-state.
Notice that this expansion implies from Eq.~\eqref{expanKG}, that the wavefunction $\psi(a\rightarrow 0,\theta)\rightarrow 0$, which is the desired behavior.
Here, $N$ is a number that can be chosen arbitrarily. The approximation improves as $N$ increases.
According to the SM, we can expand also the following functions
\begin{eqnarray}\label{SmKG2}
  V_{\mbox{\footnotesize{eff}}}(a)\,  u_n(a)&\approx& \sum_{m=1}^N B_m^{(n)} \sin\left(\frac{m\pi}{L}a\right)\, ,\nonumber\\
 \frac{9}{4 a^2} u_n(a)&\approx& \sum_{m=1}^N C_m^{(n)} \sin\left(\frac{m\pi}{L}a\right)\, ,
\end{eqnarray}
where again $B_m^{(n)}$ and $C_m^{(n)}$ are coefficients depending on the $n$-state. It is straightforward to find the relation between $B_m^{(n)}$ and $C_m^{(n)}$ with $A_m^{(n)}$. Using \eqref{SmKG} in \eqref{SmKG2}, we get that $B_m^{(n)}=\sum_{l=1}^N D_{ml} A_l^{(n)}$, and $C_m^{(n)}=\sum_{l=1}^N F_{ml} A_l^{(n)}$, where
\begin{eqnarray}
  D_{ml} &=& \frac{2}{L}\int_0^L \sin\left(\frac{m\pi}{L}a\right) V_{\mbox{\footnotesize{eff}}}(a) \sin\left(\frac{l
  \pi}{L}a\right) da\, ,\nonumber\\
  F_{ml} &=& \frac{9}{2L}\int_0^L \sin\left(\frac{m\pi}{L}a\right) \frac{1}{a^2}\sin\left(\frac{l
  \pi}{L}a\right) da\, .
\end{eqnarray}

Now, with the expansion \eqref{SmKG} and \eqref{SmKG2} in Eq.~\eqref{KGquantumechanicalEq} we can finally obtain the eigenvector equation
\begin{equation}\label{}
\mathbb{F}^{-1}\cdot\mathbb{K}\cdot\mathbb{A}^{(n)}= E_n^2 \mathbb{A}^{(n)}\, ,
\end{equation}
for the $\mathbb{A}^{(n)}$ vector (with $N$ components $A_l^{(n)}$) and where $E_n$ corresponds to the eigenvalues. Here $\mathbb{F}^{-1}$ is the $N\times N$ inverse matrix of $\mathbb{F}$ (with components $F_{ml}$), and the $N\times N$ matrix $\mathbb{K}$ has the components $D_{ml}+({m\pi}/{L})^2\delta_{ml}$.
The dimensions of the matrices $\mathbb{F}$ and $\mathbb{K}$ will be fixed once a cut on the series \eqref{SmKG} in \eqref{SmKG2} will be done. While the better the approximation, the larger will be the matrices.

Thus, the system is completed solved. We now can understand that physical content of $E_n$. Their values correspond to the energies of the different possible states of the Universe. This also allow us to identify the variable $\theta$  as a super-time, as it was previously discussed.

If one is being less restrictive with the assumption that the potential $V(\theta)$ is constant, Eq.~\eqref{KGfinalG} can be written in the form of the Wheeler-DeWitt Super-Hamiltonian formalism \cite{hawk1,hawk2}. For example, for a closed universe $k=1$ (and the  case quinteessence $\epsilon=-1$) it is possible to rewrite  Eq.~\eqref{KGfinalG} as
\begin{equation}\label{}
  \frac{\partial^2\psi}{\partial \alpha^2}-\frac{\partial^2\psi}{\partial \varphi^2}+\left(\hat m^2\varphi^2 e^{6\alpha}-e^{4\alpha}\right)\psi\equiv {\cal H}\psi =0
\end{equation}
where $\alpha=\ln a$, and $\varphi=2\theta/3$. To obtain this equation we chose $m^2=1/18$ and $V(\theta)=3\hat m^2\varphi^2$, where $\hat m$ is the mass of the  field.
Here, ${\cal H}$ is usually called the Wheeler-DeWitt  Super-Hamiltonian \cite{hawk1,hawk2}. Thus, the quantization of the FRWq system as a Klein-Gordon particle proposed here can reproduce known results of quantization using the Super--Hamiltonian formalism.

On the other hand, another possible solution of Eq.~\eqref{KGquantumechanicalEq} could be achieved using the Frobenius (polynomial) method, different from the SM. This solution corresponds to a polynomial expansion in $a$ where all the coefficients can be found from recurrence relations.  A polynomial solution for $u_n$ has the form
\begin{equation}
u_n(a)\approx a^y \sum_{m=0}^\infty b_m^{(n)} a^m\, ,
\end{equation}
where $y>0$ and $b_m^{(n)}$ are constants. For the sake of simplicity we choose $b_0^{(n)}=1$. Using the previous expansion in Eq.~\eqref{KGquantumechanicalEq} we find
\begin{eqnarray}
&&\sum_{m=0}^\infty \left[(m+y)(m+y-1)+\frac{9E_n^2+1}{4}\right] b_m^{(n)} a^{m+y-2}\nonumber\\
&&-\sum_{m=4}^\infty18 k\,  b_{m-4}^{(n)} a^{m+y-2} +\sum_{m=6}^\infty 6V\,  b_{m-6}^{(n)} a^{m+y-2}=0\, \nonumber\\
&&
\end{eqnarray}
where $V$ is defined after Eq.~\eqref{FRWqe2}.
Equating every term to zero, we can readily find the solutions
\begin{eqnarray}
y&=&\frac{1}{2}\pm\frac{3i}{2}E_n\, ,\nonumber\\
b_4^{(n)}&=&\frac{72k}{\left(3iE_n+9\right)\left(3iE_n+7\right)}\, ,\nonumber\\
b_6^{(n)}&=&\frac{-24 V}{\left(3iE_n+13\right)\left(3iE_n+11\right)}\, ,\nonumber\\
b_8^{(n)}&=&\frac{72k}{\left(3iE_n+17\right)\left(3iE_n+15\right)}\, ,
\end{eqnarray}
with also $b_2^{(n)}=0$, $b_{2m+1}^{(n)}=0$, and the recurrence relation for $m\geq 5$ is
\begin{eqnarray}
b_{2m}^{(n)}=\frac{72k b_{2m-4}^{(n)}-24 V b_{2m-6}^{(n)} }{\left(3iE_n+4m+1\right)\left(3iE_n+4m-1\right)}\, ,
\end{eqnarray}
where now the problem is complety solved.
The weakness of this method is that it does not give any physical meaning to the constant $E_n$ in the ansatz \eqref{expanKG}.

\subsection{Quantization of the FRWq system {\it a la} Dirac-Pauli}
\label{quantiDira}

Basically, the quantization process proposed here consists in finding the square root of the Hamiltonian operator \eqref{momentumoperator1}. In principle one can use matrices to find the square root, but its use implies the notion that the Universe behaves as a Dirac particle.

At a first glance it may appear strange to quantize a model for a relativistic pointlike particle with a quantum spin theory. However, in 1928, Breit \cite{breit}  showed that there exist a correspondence between the Dirac and the relativistic pointlike particle Hamiltonians. In that work, it is shown that one can obtain the Dirac equation via a  prescription of replacement of the particle velocity and the Dirac matrices (as well as the prescription in Schr\"odinger or Klein-Gordon theories where the energy and  momentum and  can be replaced by the time and space derivatives). Thus, the Breit's prescription implies a classical and geometrical interpretation of the spin. We leave the calculations and the deep discussion of this idea to Sec.~\ref{discus}. For now, in this section we restrict ourselves  to  follow Breit's interepretation and to perform the quantization of the FRWq Universe using Dirac matrices.

We propose that the Hamiltonian \eqref{momentumoperator1} can be written using Dirac matrices ($\alpha$ and $\beta$). This will give us the Hamiltonian operator
\begin{equation}\label{momentumoperatorDirac}
  {\cal H}=\Omega^{1/2} \left(\alpha\cdot\hat p+\beta\right) \Omega^{1/2}\, .
\end{equation}
In Sec.~\ref{discus} we justify this choice.
This Hamiltonian allows us to quantize a FRWq Universe as if it were a relativistic spin particle. Using the operator \eqref{momentumoperatorDirac}, the quantum mechanical equation \eqref{quantumeq} now reads
\begin{equation}\label{eqCuanD}
  i\frac{\partial\Psi}{\partial \theta}=-i\alpha^\xi\left[\frac{\partial}{\partial\xi}+\frac{1}{2}\frac{\partial\ln\Omega}{\partial\xi}\right]\Psi+\beta\Omega\Psi\, ,
\end{equation}
where $\Psi$ now is a bi-spinor. Here, $\alpha^\xi$ stands for any of the Dirac matrices $\alpha$. Choosing the  Dirac representation $\gamma^0=\beta$ and $\gamma^\xi=\gamma^0\alpha^\xi$, we can operate the Eq.~\eqref{eqCuanD} by $\gamma^0$ by the left to find that
\begin{equation}\label{diracQQQ}
  i\gamma^0\frac{\partial\Psi}{\partial\theta}+i\gamma^\xi\left(\frac{\partial}{\partial\xi}+\frac{1}{2}\frac{\partial\ln\Omega}{\partial\xi}\right)\Psi=\Omega\Psi\, .
\end{equation}

The previous equation describes the quantum theory for the FRWq Universe modeled as a Dirac particle. Also, this equation can be obtained directly from the theory of the Dirac equation in curved spacetimes, thus giving validity to our quantization scheme (shown in the Appendix \ref{DiracEcst}).

Dirac matrices are $4\times 4$ and as we are describing a two-dimensional conformal system, we may anticipate that the above equation is reducible.  This means that Dirac equation \eqref{diracQQQ} couples the wavefunction in pairs, implying that the the two pairs of wavefuntions satisfy the same equation. This gives us a hint that a completely equivalent quantization formalism to the previous one can be achieved using Pauli matrices. Solving the Hamiltonian \eqref{momentumoperator1}  using Pauli matrices (notice that there is no restriction to this ansatz), the Hamiltonian operator \eqref{momentumoperator1} can be written as
\begin{equation}\label{momentumoperatorPauli}
  \mathcal H=\Omega^{1/2}\left(\sigma_x+\sigma_y {\hat p} \right) \Omega^{1/2}\, ,
\end{equation}
where $\sigma_x$ and $\sigma_y$ can be any two different Pauli matrices. The quantum mechanical equation \eqref{quantumeq} describing the FRWq system becomes (setting $\hbar=1$)
\begin{equation}\label{quantumeqPauli}
  i\mathbf 1\frac{\partial\Psi}{\partial\theta}= \Omega\, \sigma_x \Psi-i\sigma_y \left[\frac{\partial}{\partial\xi}+\frac{1}{2}\frac{\partial\ln\Omega}{\partial\xi}\right]\Psi\, ,
\end{equation}
where $\mathbf 1$ is the unit matrix, and $\Psi$ represents a spinor field.
We can use the Pauli matrices properties to put the previous equation in the following form
\begin{equation}\label{quantumeqPauli2}
  i\sigma_x\frac{\partial\Psi}{\partial\theta}-\sigma_z \left[\frac{\partial}{\partial\xi}+\frac{1}{2}\frac{\partial\ln\Omega}{\partial\xi}\right]\Psi= \Omega\, \Psi\, ,
\end{equation}
where $i\sigma_z=\sigma_x\sigma_y$. It can be proves that choosing  $\sigma_x=\sigma_3$ and $\sigma_z=\sigma_2$, gives the same dynamical equation than choose $\gamma^\xi=\gamma^1$  in the Dirac equation.

On the other hand, from Eq.~\eqref{diracQQQ}  we notice a that defining the bi-spinor $\Psi'=\sqrt{\Omega}\Psi$,  we can obtain the equation
\begin{equation}\label{diracQQQpotential}
 i\left(\gamma^0\frac{\partial}{\partial\theta}+\gamma^\xi\frac{\partial}{\partial\xi}\right)\Psi'=\Omega\Psi'\, ,
\end{equation}
which is a flat 1+1 spacetime massless Dirac equation with a scalar potential. This kind of equation have been extensively studied and approximated solutions have been found \cite{sanchez,yesil,eleuch,cooper}.

Finally, using the wavefunction $\Psi$ (given by solving either the Dirac or Pauli equations), we can calculate the probability density of the Dirac field as
\begin{equation}\label{}
  |\Psi(\theta)|^2=\int_0^{a_f} \frac{da}{a}\Omega\, \Psi^\dag \Psi\, ,
\end{equation}
where $\Psi^\dag$ is the transpose conjugated of the wavefunction $\Psi$. In the previous expression $\Psi$ and $\Psi^\dag$ are written in terms of $a$.  In a similar fashion, the expected value of the scale factor for the Universe under the Dirac quantization is obtained
\begin{equation}\label{valoraDirac}
\left\langle a(\theta)\right\rangle=\frac{1}{|\Psi|^2}\int_0^{a_f} da\,  \Omega\,  \Psi^\dag \Psi\, ,
\end{equation}
depending again on the values of super-time $\theta$.

To further study the system let us do the bi-spinor descomposition
\begin{equation}\label{wavefunctionDi}
  \Psi(\theta,\xi)=\sum_{n=0}^\infty e^{i E_n\theta}\left(
                                 \begin{array}{c}
                                   \psi_n (\xi)\\
                                   \zeta_n (\xi)\\
                                   \varphi_n(\xi) \\
                                   \chi_n(\xi) \\
                                 \end{array}
                               \right)\, ,
       \end{equation}
where $E_n$ are constants. Using \eqref{wavefunctionDi} in \eqref{diracQQQ}, gives
\begin{eqnarray}\label{DiracQex1}
0&=&\frac{d\chi_n}{d\xi}+\frac{1}{2\Omega}\frac{d\Omega}{d\xi}\chi_n+i (E_n+\Omega)\psi_n\, ,\nonumber\\
  0&=&\frac{d\psi_n}{d\xi}+\frac{1}{2\Omega}\frac{d\Omega}{d\xi}\psi_n+i (E_n-\Omega)\chi_n\, ,
\end{eqnarray}
where we have made the particular choice of the Dirac matrix
\begin{equation}\label{}
  \gamma^\xi=\gamma^1=\left(
                        \begin{array}{cccc}
                          0 & 0 & 0 & 1 \\
                          0 & 0 & 1 & 0 \\
                          0 & -1 & 0 & 0 \\
                          -1 & 0 & 0 & 0 \\
                        \end{array}
                      \right)\, .
\end{equation}
The same mathematical equations can be obtained for the fields $\zeta_n$ and $\varphi_n$, reflecting that the system can also be studied using Pauli matrices.
Now the fields $\psi_n$ and $\chi_n$ appear coupled.
From those equations it is not possible to recover the Klein-Gordon equation \eqref{KGquantumechanicalEq}. The reason is that wavefunctions are coupled to the spacetime metric through the potential due to the quintessence field.

First, let us notice that a simple  exact solution of Eq.~\eqref{DiracQex1} can be found when the fields do not depend on quintessence, i.e. when $E_n=0$. In this case, the solutions are
\begin{eqnarray}
\psi_n(\xi)= \frac{i}{\sqrt\Omega}\exp\left(\Omega_I\right)\, ,\quad \chi_n(\xi)= \frac{1}{\sqrt\Omega}\exp\left(\Omega_I\right)\, ,
\end{eqnarray}
where we define
\begin{equation}\label{OmegaI}
\Omega_I(\xi)\equiv\left(\frac{1}{2}-2\frac{3^{1/3}k}{V}e^{-4\xi/3}\right)\sqrt{\frac{3^{4/3} k}{2}e^{8\xi/3}-\frac{3Ve^{4\xi}}{8}}\, .
\end{equation}
However, these solutions are not well-behaved at $\xi\rightarrow -\infty$ ($a\rightarrow 0$) as it diverges. Therefore we will seek for solutions with $E_n\neq 0$.

A more general solution can be obtained in the following way. Let us define $\phi^+_n(\xi)=\sqrt{\Omega}\, \psi_n(\xi)$, and $\phi^-_n(\xi)=\sqrt{\Omega}\, \chi_n (\xi)$. Thus, the Eqs.~\eqref{DiracQex1} can be re-written as
\begin{equation}
\frac{d\phi_n^{\pm}}{d\xi}=\Omega_\mp \phi_n^\mp\, ,
\end{equation}
where $\Omega_\pm=\pm i \left(E_n\pm\Omega\right)$, such that $\Omega_+\Omega_-=E_n^2-\Omega^2$. The previous equation are coupled, but we can find the following second-order equation which holds for each of the fields
\begin{equation}
\left( -\frac{d^2}{d\xi^2}+ \frac{1}{\Omega_\mp}\frac{d\Omega^{\mp}}{d\xi}\frac{d}{d\xi}   +\Omega_+\Omega_-\right)\phi_n^\pm=0\, .
\end{equation}

The above equation can be reduced to familiar expressions doing the change $\phi_n^\pm=\sqrt{\Omega_\mp}\exp\left(i\int \mu_\pm(\xi') d\xi' \right)$. The equation for $\mu$ is reduced to a Ricatti equation
\begin{equation}\label{ricattiDirac}
\frac{d\mu_\pm}{d\xi}+i\mu_\pm^2-iV_\pm=0\, ,
\end{equation}
with
\begin{equation}
V_\pm=-\frac{3}{4\Omega_\mp^2}\left(\frac{d\Omega_\mp}{d\xi}\right)^2+\frac{1}{2\Omega_\mp}\frac{d^2\Omega_\mp}{d\xi^2}-\Omega_-\Omega_+\, .
\end{equation}

A general solution of the Ricatti equation ~\eqref{ricattiDirac} can be found if we are able to find a particular solution.

On the other hand, when the solutions depend on the quintessence-tlaplon field, we can use the SM to completely solve the Dirac equations \eqref{DiracQex1}, as well as in the previous section. As it was shown before, this method allows us to reduce the complicated equations \eqref{DiracQex1} to an eigenvalue problem for any $k$. First, we define the wavefunctions $u_n(a)=\sqrt{\Omega}\, \psi_n (a)/a$ and
$v_n(a)=\sqrt{\Omega}\, \chi_n (a)/a$, where now $\psi_n$ and $\chi_n$ should be written in terms of $a$ (similar changes can be done for the fields $\zeta_n$ and $\varphi_n$). These two new functions satisfy the equations
\begin{eqnarray}\label{Diracu}
  0 &=& a\frac{d v_n}{da}+v_n+i\left(E_n+\Omega\right) u_n\, ,\nonumber \\
  0 &=& a\frac{d u_n}{da}+u_n+i\left(E_n-\Omega\right) v_n\, .
\end{eqnarray}

We can now use the SM for the functions $u_n$ and $v_n$ (notice that every term in Eqs.~\eqref{Diracu} approach to zero as $a$ goes to zero). With the SM, we can assume the following dependence for the different functions
\begin{eqnarray}\label{SmD}
  u_n(a)&\approx& \sum_{m=1}^N A_m^{(n)} \sin\left(\frac{m\pi}{L}a\right)\, ,\nonumber\\
 v_n(a)&\approx& \sum_{m=1}^N B_m^{(n)} \sin\left(\frac{m\pi}{L}a\right)\, ,
\end{eqnarray}
where again  $A_m^{(n)}$ and  $B_m^{(n)}$ are constant coefficients that depend on the $n$-state. Anew, the wavefunction \eqref{wavefunctionDi} behaves as $\Psi(\theta, a\rightarrow 0)\rightarrow 0$.
Similarly, we get
\begin{eqnarray}\label{SmD2}
  \Omega(a)\,  u_n(a)&\approx& \sum_{m=1}^N C_m^{(n)} \sin\left(\frac{m\pi}{L}a\right)\, ,\nonumber\\
  \Omega(a)\,  v_n(a)&\approx& \sum_{m=1}^N D_m^{(n)} \sin\left(\frac{m\pi}{L}a\right)\, ,\nonumber\\
a \frac{d u_n(a)}{da}&\approx& \sum_{m=1}^N E_m^{(n)} \sin\left(\frac{m\pi}{L}a\right)\, ,\nonumber\\
a \frac{d v_n(a)}{da}&\approx& \sum_{m=1}^N F_m^{(n)} \sin\left(\frac{m\pi}{L}a\right)\, ,
\end{eqnarray}
where the relations between the coefficients are $C_l^{(n)}=\sum_{l=1}^N G_{lm} A_m^{(n)}$,  $D_l^{(n)}=\sum_{l=1}^N G_{lm} B_m^{(n)}$, $E_l^{(n)}=\sum_{l=1}^N H_{lm} A_m^{(n)}$ and $F_l^{(n)}=\sum_{l=1}^N H_{lm} B_m^{(n)}$, with the matrix elements
\begin{eqnarray}\label{GmlHml}
  G_{lm} &=& \frac{2}{L}\int_0^L \sin\left(\frac{l\pi}{L}a\right) \Omega(a) \sin\left(\frac{m  \pi}{L}a\right) da\, ,\nonumber\\
  H_{lm} &=& \frac{2 m \pi}{L^2}\int_0^L a\sin\left(\frac{l\pi}{L}a\right)\cos\left(\frac{m  \pi}{L}a\right) da\, .
\end{eqnarray}

Using Eqs.~\eqref{SmD} in Eqs.~\eqref{SmD2}, and the previous relations on Eqs.~\eqref{Diracu},  we find after some algebra the eigenvector equation
\begin{equation}\label{eigenD}
\mathbb{M}\cdot\mathbb{V}^{(n)}= E_n \mathbb{V}^{(n)}\, ,
\end{equation}
where $E_n$ corresponds to the eigenvalues (associated to the energy) and
the eigenvector $\mathbb{V}^{(n)}$ is formed by the $2N$ components  $A_m^{(n)}$ and  $B_m^{(n)}$ as
\begin{equation}\label{vectorDiraceigen}
\mathbb{V}^{(n)}=\left(
                                 \begin{array}{c}
                                    A^{(n)}\\
                                   B^{(n)}
			\end{array}
                               \right)\, ,
       \end{equation}
and the $2N\times 2N$ matrix $\mathbb{M}$ is such that
\begin{equation}
\mathbb{M}=\left(
                                 \begin{array}{cc}
                                   -\mathbb{G}&   i \left(\mathbb{I}+\mathbb{H} \right)\\
                                     i \left(\mathbb{I}+\mathbb{H} \right)&  \mathbb{G}
			\end{array}
                               \right)\, ,
       \end{equation}
where the $N\times N$ matrices $\mathbb{G}$ and $\mathbb{H}$ are constructed by the components given in Eqs.~\eqref{GmlHml} ($\mathbb{I}$ is the identity matrix).

The evolution of the system is completely determined by solving the eigenvector and eigenvalue equation \eqref{eigenD}.   The approximated solution improves by increasing $N$.

\subsection{Quantization of the FRWq system  {\it a la} Majorana}
\label{quantiMajo}

Strictly speaking, the Dirac description of the FRQW system implies that the Universe can interact with  self-electromagnetic fields. One way to avoid this issue is to use Majorana matrices instead of Dirac matrices. In this way the quantization scheme models a Universe with quintessence as a neutral relativistic quantum particle. The quantum mechanical equation is
\begin{equation}\label{diracQQQM}
  i\gamma^0_M\frac{\partial\Psi_M}{\partial\theta}+i\gamma^\xi_M \left(\frac{\partial}{\partial\xi}+\frac{1}{2}\frac{\partial\ln\Omega}{\partial\xi}\right)\Psi_M=\Omega\Psi_M\, .
\end{equation}
where now $\gamma_M$ are the Majorana matrices, and $\Psi_M$ represents the wavefunction of the FRWq system in the Majorana scheme of quantization.

Similarly to the previous case, the expected value of the scale factor for the Universe under the Majorana quantization is
\begin{equation}\label{valoraMajo}
\left\langle a(\theta)\right\rangle=\frac{1}{|\Psi_M|^2}\int_0^{a_f} da\,  \Omega\,  \Psi_M^\dag \Psi_M\, ,
\end{equation}
where $\Psi_M^\dag$ is the transpose conjugated of the wavefunction $\Psi_M$, and the probability density is $|\Psi_M|^2=\int_0^{a_f} {da}\, \Omega\, \Psi_M^\dag \Psi_M/a$. As well as all the previous cases, the expected value of the scale factor depends on $\theta$.

Analytical representation of the solutions can be obtained by performing the descomposition
\begin{equation}\label{wavefMajo}
  \Psi_M(\theta,\xi)=\sum_{n=0}^\infty e^{i E_n\theta}\left(
                                 \begin{array}{c}
                                   \psi_n (\xi)\\
                                   \zeta_n (\xi)\\
                                   \varphi_n(\xi) \\
                                   \chi_n(\xi) \\
                                 \end{array}
                               \right)\, ,
       \end{equation}
and choosing
\begin{equation}\label{}
      \gamma^\xi_M=\gamma^1_M=\left(
                        \begin{array}{cccc}
                          i & 0 & 0 & 0 \\
                          0 & -i & 0 & 0 \\
                          0 & 0 & i & 0 \\
                          0 & 0 & 0 & -i \\
                        \end{array}
                      \right)\, ,
\end{equation}
Eq.~\eqref{diracQQQM} may be rewritten as
\begin{eqnarray}
  0&=&\frac{d \psi_n}{d \xi}+\left(\frac{1}{2\Omega}\frac{d \Omega}{d \xi}+\Omega\right)\psi_n -iE_n\chi_n\, ,\nonumber\\
    0&=&\frac{d \chi_n}{d \xi}+\left(\frac{1}{2\Omega}\frac{d \Omega}{d \xi}-\Omega\right)\chi_n -iE_n\psi_n\, .\label{DiracQex4M}
  \end{eqnarray}
Again the wavefunction $\psi_n$ is coupled to $\chi_n$. Similarly $\zeta_n$ and $\varphi_n$ are coupled by the same mathematical equations.

Notice that unlike the Dirac case, when $E_n=0$, the equations in the Majorana scheme decouple. In this case, the simple solutions can be obtained as
\begin{eqnarray}
\psi_n(\xi)= \frac{1}{\sqrt\Omega}\exp\left(-\Omega_I\right)\, ,\quad \chi_n(\xi)= \frac{1}{\sqrt\Omega}\exp\left(\Omega_I\right)\, ,
\end{eqnarray}
where we use the definition \eqref{OmegaI} for $\Omega_I$. As in Dirac scheme, these solutions again diverge for $\xi\rightarrow -\infty$ ($a\rightarrow 0$). For $E_n\neq 0$ we can find a modified Klein-Gordon equation for $\psi_n$
\begin{eqnarray}\label{majoranapsi}
 0&=& \frac{d^2\psi_n}{d\xi^2}+\frac{1}{\Omega}\frac{d\Omega}{d\xi}\frac{d \psi_n}{d\xi}\nonumber\\
&&+\left(\frac{1}{2\Omega}\frac{d^2\Omega}{d\xi^2}-\frac{1}{4\Omega^2}\left(\frac{d\Omega}{d\xi}\right)^2+\frac{d\Omega}{d\xi}-\Omega^2+E_n^2\right)\psi_n\, ,\nonumber\\
&&
\end{eqnarray}
which contains an effective mass term depending on curvature variations [compare with Eq.~\eqref{KGquantumechanicalEq}].
For the special case of $k=0$, this equation reduces to the equation for a diatomic molecule decribed by the Morse potential \cite{sahfaa}.  Performing the change of variables $\psi=\exp(-\xi)\, \varphi$,  Eq.~\eqref{majoranapsi} simplify to
\begin{equation}\label{majoranapsik0}
\frac{\partial^2\varphi}{\partial\xi^2}=\left(-\frac{3}{8}V e^{4\xi}-E^2-\sqrt{-\frac{3}{2}V} e^{2\xi}\right)\varphi\, .
\end{equation}
As it was discussed in the begining of Sec.~\ref{quantizationquantization}, the general potential $\bar V$ must be positive. Therefore we can choose a representation of $V=V(\theta)=-(8/3)\exp(2x_e)$, where $x_e$ is a constant. Making another change of variables $x=-2\xi$ and defining ${\cal E}=E^2/4$, Eq.~\eqref{majoranapsik0} can be put in the form
\begin{equation}\label{majoranaMorse}
{\cal E}\varphi=\left[-\frac{\partial^2}{\partial x^2}+\frac{1}{4}\left(e^{-2(x-x_e)}-2e^{-(x-x_e)}\right)\right]\varphi\, ,
\end{equation}
which is the quantum equation for a diatomic molecule described by the Morse potential.
The wavefunctions and energy spectrum of this problem are already known.

Another important feature of Eqs.~\eqref{DiracQex4M} deserves to be highlighted. Defining the new wavefunctions $\phi^+_n=\sqrt{\Omega}\psi_n$ and $\phi^-_n=\sqrt{\Omega}\chi_n$, then Eqs.~\eqref{DiracQex4M} can be re-expressed as
\begin{equation}\label{susyMajo}
Z_\pm\phi_n^\pm=\pm i E_n\phi_n^\mp\, ,
\end{equation}
where the operators are defined as \cite{sahfaa}
\begin{equation}\label{operatorsusyMajo}
Z_\pm=\pm\frac{d}{d_\xi}+\Omega\, .
\end{equation}

Notice that \eqref{susyMajo} represent the equations for supersymmetric quantum mechanics \cite{cooper,crom,cooper2}. Each wavefunction satisfies
\begin{equation}
H_1\phi_n^+=\frac{E_n^2}{2}\phi_n^+\, ,\qquad H_2\phi_n^-=\frac{E_n^2}{2}\phi_n^-\, ,
\end{equation}
where the Hamiltonians
\begin{eqnarray}
H_1&=&\frac{1}{2}Q_-Q_+=-\frac{1}{2}\frac{d^2}{d\xi^2}+\frac{1}{2}\left(-\frac{d\Omega}{d\xi}+\Omega^2\right)\, ,\nonumber\\
H_2&=&\frac{1}{2}Q_+Q_-=-\frac{1}{2}\frac{d^2}{d\xi^2}+\frac{1}{2}\left(\frac{d\Omega}{d\xi}+\Omega^2\right)\, ,
\end{eqnarray}
can be used to define the Super-Hamiltonian
\begin{equation}
H_M=\left(\begin{array}{cc}
                          H_1 & 0\\
                          0 & H_2\\
                        \end{array}
                      \right)\, ,
\end{equation}
used to write the above equations as
\begin{equation}\label{}
H_M\left( \begin{array}{c}
                          \phi_n^+ \\
\phi_n^- \\
                        \end{array}
                      \right)=\frac{E_n^2}{2}\left( \begin{array}{c}
                          \phi_n^+ \\
\phi_n^- \\
                        \end{array}
                      \right)\, .
\end{equation}

The operators \eqref{operatorsusyMajo} can be used to define the supercharges
\begin{equation}
Q=\left(\begin{array}{cc}
                          0 & 0\\
                          Z_+ & 0\\
                        \end{array}
                      \right)\, ,\qquad
Q^\dag=\left(\begin{array}{cc}
                          0 & Z_-\\
                          0 & 0\\
                        \end{array}
                      \right)\, ,
\end{equation}
which are operators that can change bosonic (fermionic) states into fermionic (bosonic) ones. The above supersymmetric system exhibits the same features of any other supersymmetric quantum theory \cite{cooper2}.

Similar to previous sections, analytical approximated solutions for any $k$ can be found using the SM approach. As before, we can perform this task defining the variables $u_n=\sqrt\Omega\, \psi_n/a$ and $v_n=\sqrt\Omega\, \chi_n/a$, that satisfy the following equations
\begin{eqnarray}\label{uvnMajor}
  0&=&a\frac{d u_n}{d a}+\left(1+\Omega\right)u_n -iE_n v_n\, ,\nonumber\\
 0&=&a\frac{d v_n}{d a}+\left(1-\Omega\right)v_n -iE_n u_n\, .\label{DiracQex4M2}
  \end{eqnarray}
Again notice that the SM allows us to have a well-defined behavior of the wavefunction \eqref{wavefMajo}  as $\Psi_M(\theta, a\rightarrow 0)\rightarrow 0$.

Applying the SM means we have to use similar decompositions \eqref{SmD}, \eqref{SmD2} and \eqref{GmlHml} to the Majorana case. For simplicity we use the same notation as before. Eqs.~\eqref{uvnMajor} can be finally written as
\begin{equation}\label{eigenM}
\mathbb{K}\cdot\mathbb{V}^{(n)}= E_n \mathbb{V}^{(n)}\, ,
\end{equation}
where again $E_n$ are the eigenvalues and  $\mathbb{V}^{(n)}$ is the vector \eqref{vectorDiraceigen}. The $2N\times 2N$ matrix $\mathbb{K}$ is now
\begin{equation}
\mathbb{K}=-i\left(
                                 \begin{array}{cc}
                                  \mathbb{O}&   \mathbb{I}+\mathbb{H}-\mathbb{G} \\
                                      \mathbb{I}+\mathbb{H}+\mathbb{G}&  \mathbb{O}
			\end{array}
                               \right)\, ,
       \end{equation}
where again  $\mathbb{G}$ and $\mathbb{H}$ are $N\times N$ matrices constructed by Eqs.~\eqref{GmlHml}, and $\mathbb{O}$ is the $N\times N$ zero matrix.

\section{Discussions}
\label{discus}

The quantization schemes presented before (Klein-Gordon, Dirac or Majorana) are only possible  due the unification between the FRW geometry and the quintessence scalar field (tlaplon field \cite{hrrs}) with a Fermat-like Lagrangian for a relativistic particle moving on conformally flat spacetime.

The quantum equations obtained for every case can be considered as generalizations of the Wheeler-DeWitt Super-Hamiltonian formalism.
Our proposal establishes that the quintessence-tlaplon field could be necessary as a first step to construct a  geometrically unified theory for the quantization of an expanding universe (with a quintessence type of dark energy).

To our knowledge this is the first time that an attempt to construct a quantum theory for the FRW Universe is carried out using a relativistic quantum theory for Dirac particles. As it was mentioned before, the reason to perform this quantization in a relativistic pointlike particle model is that we want to follow Breit's interpretation \cite{breit}. Breit showed that the spin can emerge as a geometrical and dynamical interpretation of the classical velocity of the particle. It was claimed that the Dirac matrices follow a replacement prescription similar to  those largely used in non-relativistic quantum mechanics and spinless relativistic quantum mechanics. He showed this for a particle on flat spacetime. Here we will follow his prescription for a relativistic particle in a conformally two-dimensional flat space.

From the classical Hamiltonian \eqref{Hclass1} we can obtain the relation $H^2=g_{00}+\pi^2$, where we have made use of $g_{00}=\Omega^2$. From here we can obtain that
\begin{equation}
\frac{\sqrt{g_{00}}}{H}=\sqrt{1-q^2}\, ,
\end{equation}
where we have defined the velocity variable
\begin{equation}
q=\frac{\pi}{H}\, .
\end{equation}

Using this variable, the  Hamiltonian \eqref{Hclass1} can be re-written as $H={g_{00}}/{H}+{\pi^2}/{H}$, or better
\begin{equation}
H=\sqrt{g_{00}}\left(\frac{q\pi}{\Omega}+\sqrt{1-q^2}\right)\, .
\end{equation}

Breit's interpretation corresponds to the identification of the Dirac matrices as \cite{breit}
\begin{equation}\label{identificationa}
q\rightarrow \alpha\, ,\qquad \sqrt{1-q^2}\rightarrow\beta\, ,
\end{equation}
allow us to construct  the quantum Hamitonian  \eqref{momentumoperatorDirac}  (with the definition \eqref{operDD} for momentum operator) from its classical analogue  \eqref{Hclass1}.
Breit showed that the identification \eqref{identificationa} is consistent with the postulates of Dirac theory. The implications of this prescription have started to be investigated with the purpose of to understand the underlying nature of the spin or antiparticles \cite{savasta}.
Therefore, the previous interpretation gives validity to the quantum theories developed in Secs.~\ref{quantiDira} and \ref{quantiMajo}, and also it is shown in Appendix C that both treatments are completely equivalent to the well-established Dirac theory in curved spacetimes.

It is important to discuss a common concept appearing in the three different quantum theories of Sec.\ref{quantizationquantization}. We can identify a new time variable $\theta$ for the quantized Universe, suggesting that the Universe evolves along this super-time, not with the usual time coordinate. It is worth noting that the super-time $\theta$ is actually the quintessence-tlaplon field $\phi=\sqrt{8/3}\, \theta$ by Eq.~\eqref{theta}, which implies that the evolution of a quantum Universe is not possible in the absence of quintessence or torsion.  This could be thought as a possible origin for the dark energy.
An argument can be given in the following way. As the quantum Universe evolves  through to the super-time, we classically detect that quantum evolution as a quintessence field wich produces a cosmological negative pressure as dark energy. This also suggests that the nature of dark energy is closely related to a cosmological and dynamical  torsion of the spacetime.

Another interesting feature of the quantization theories of Secs.~\ref{quantiDira} and \ref{quantiMajo} is the physical meaning of the wavefunctions components of the bi-spinor. In comparison with the Dirac theory for physical particles, we can recognize that every component of the wavefunctions introduced along this work has the dynamics of an identity in interaction with the others. Hence, we can argue that each component represents a Universe, evidencing that the Dirac bi-spinor wavefunction \eqref{wavefunctionDi}, or the Majorana bi-spinor wavefunction \eqref{wavefMajo}, represents a description of a Multiverse.
The different Universes $\psi$ and $\chi$ (or $\zeta$ and $\varphi$) appear coupled, implying an interaction between the Universes. In this way, the Multiverse behaves as a di-atomic molecule. The interaction between the wavefunction of the Universes produces the dynamical evolution of the expected value of the scale factor, as it can be seen from Eqs.~\eqref{valoraDirac} and \eqref{valoraMajo}. Therefore the expansion rate of the scale factor, produced by quintessence (dark energy),  could be a direct consequence of the existence of a Multiverse. This idea will be explored in forthcoming works.
Moreover, this kind of Multiverse theory is different from the Everett's Many-worlds interpretation of quantum mechanics \cite{everett1,everett2}. Every  Universe, characterized by the wavefunction $\psi$ and $\chi$ (or $\zeta$ and $\varphi$) can be interpreted using Everett's theory, but they are part of a larger duality expressed by the bi-spinor. The Many-worlds interpretation of Multiverses has been largely studied (the literature is vast and some few examples are in Refs.~\cite{weinberg,tegmark,tegmark2,tegmark3,feeney,hall,antonovaa,vilekin,carroll1,carroll2,carroll3}), and therefore we believe this work could be a contribution along these lines.
Another important point to be highlighted is that according to Eqs.~\eqref{susyMajo} we can infer that the Multiverse
has the structure of a supersymmetric system, being the two Universes the super-partners of each other. We can use all the well-known tools of supersymmetry \cite{cooper2} to study the main characteristics of this Mutliverse theory.  This new concept creates a new unifying frame for a Multiverse theories and it deserves to be deeply explored in the future.

Finally, we would like to remark that the three quantization schemes are only possible due to the geometrical unification between spacetime and the quintessence-tlaplon field  proposed in Secs.~II and III. In the case of the Klein-Gordon theory, the unification leads to results that coincide with similar ones in literature. We expect that this unification can bring new insights in the field of quantum cosmology.

\begin{acknowledgments}

The author C.R. was supported by Conicyt PhD fellowship No 21150314.

\end{acknowledgments}

\appendix
\section{Dirac equation in curved spacetimes}
\label{DiracEcst}

The curved spacetime Dirac equation is
\begin{equation}\label{}
  i {e^\mu}_d \gamma^d \left(\partial_\mu+\frac{1}{8}\omega_{ab\mu}[\gamma^a,\gamma^b]\right)\Psi=\Psi\, ,
\end{equation}
where we defined the vierbein as
\begin{equation}\label{}
  g_{\mu\nu}={e_\mu}^a {e_\nu}^b \eta_{ab}\, ,
\end{equation}
with the flat-spacetime metric $\eta_{ab}$. Notice that ${e^\mu}_d$ is the inverse vierbein in the sense that ${e_\mu}^a{e^\mu}_b=\delta^a_b$. Also we define the spin connection $\omega_{ab\mu}=\eta_{ac}{\omega^c}_{b\mu}$, with
\begin{equation}\label{}
  {\omega^c}_{b\mu}={e^c}_\nu {e^\nu}_{b,\mu}+{e^c}_\nu{e^\sigma}_b{\Gamma^\nu}_{\sigma\mu}\, ,
\end{equation}
where ${\Gamma^\nu}_{\sigma\mu}$ are the Christoffel symbols.
Because the antisymmetry of the spin connection in its first two indices, we have $\omega_{ab\mu}[\gamma^a,\gamma^b]=2\omega_{ab\mu}\gamma^a\gamma^b$.

In our two-dimensional case of the FRWq system, the vierbeins are
\begin{eqnarray}\label{}
  {e_0}^0=\Omega\, ,\,\,\, {e_1}^1=\Omega\, ,\,\,\, {e^0}_0=\frac{1}{\Omega}\, ,\,\,\, {e^1}_1=\frac{1}{\Omega}\, ,
\end{eqnarray}
Thus, the Dirac equation becomes
\begin{equation}\label{}
  i\gamma^0\left(\partial_0+\frac{1}{4}\omega_{ab0}\gamma^a\gamma^b\right)\Psi+i\gamma^1\left(\partial_\xi+\frac{1}{4}\omega_{ab1}\gamma^a\gamma^b\right)\Psi=\Omega\Psi\, .
\end{equation}
Also we have that $\omega_{ab0}\gamma^a\gamma^b=2\omega_{010}\gamma^0\gamma^1$, $\omega_{ab1}\gamma^a\gamma^b=2\omega_{011}\gamma^0\gamma^1$, and
\begin{equation}\label{}
  \omega_{010}={\Gamma^0}_{10}=\partial_\xi\ln\Omega\, , \qquad \omega_{011}={\Gamma^0}_{11}=0\, ,
\end{equation}
for the two-dimensional conformally flat metric.

Then, the Dirac equation is written as
\begin{equation}\label{}
  i\gamma^0\left(\partial_0+\frac{1}{2}\frac{\partial\ln\Omega}{\partial\xi}\gamma^0\gamma^1\right)\Psi+i\gamma^1\partial_\xi\Psi=\Omega\Psi\, ,
\end{equation}
or
\begin{equation}\label{}
  i\gamma^0\partial_0\Psi+i\gamma^1\left(\partial_\xi+\frac{1}{2}\frac{\partial\ln\Omega}{\partial\xi}\right)\Psi=\Omega\Psi\, ,
\end{equation}
which is exactly the same equation than \eqref{diracQQQ}. It is important to stress that we obtain Eq.~\eqref{diracQQQ} following a quantization scheme (with $\partial_0=\partial_\theta$).

\end{document}